\documentstyle[aps,multicol,epsf]{revtex}

\begin{document}
\draft

\title{
Giant strongly connected component of directed networks
}

\author{
S.N. Dorogovtsev$^{1, 2, \ast
}$, J.F.F. Mendes$^{1,\dagger}$, and A.N. Samukhin$^{2, \ddag
}$ 
}

\address{
$^{1}$ Departamento de F\'\i sica and Centro de F\'\i sica do Porto, Faculdade 
de Ci\^encias, 
Universidade do Porto\\
Rua do Campo Alegre 687, 4169-007 Porto, Portugal\\
$^{2}$ A.F. Ioffe Physico-Technical Institute, 194021 St. Petersburg, Russia 
}

\maketitle

\begin{abstract}
We describe how to calculate the sizes of all 
%%the 
giant connected components of a directed graph, including the {\em strongly} connected one. Just to the class of directed networks, in particular, belongs the World Wide Web. The results are obtained for graphs with statistically uncorrelated vertices and an arbitrary joint in,out-degree distribution $P(k_i,k_o)$. We show that if $P(k_i,k_o)$ does not factorize, 
the relative size of the giant strongly connected component deviates from the product of the relative sizes of the giant in- and out-components. The calculations of the relative sizes of all the giant components are demonstrated using the simplest examples. 
%%The stability of the giant strongly connected component 
%%to random damage is discussed. 
We explain that the giant strongly connected component may be less resilient to random damage than the giant weakly connected one. 
\end{abstract}


\begin{multicols}{2}

\narrowtext

%%%%%%%%%%%%%%%%%%%%%%%%%%%%%%%%%%%%%%%%%%%%%%%%%%%%%%%%%%%%%%%%%%%%%%%

The giant components of a network are components which relative sizes are finite (nonzero) in the large network limit. The knowledge of these sizes provides the basic information about the global topology of a network. The understanding of the topological structure of networks and its change under external action is the central problem of the statistical physics of random networks \cite{ajb99,ba99,ajb00,ab00a,bkm00,nsw00,cnsw00,cebh00}. 
Actually, this is 
%a part of 
the natural generalization of 
the general percolation theory. 

\begin{figure}
\epsfxsize=90mm
%%%\epsffile{fig_strong_1.eps}
\epsffile{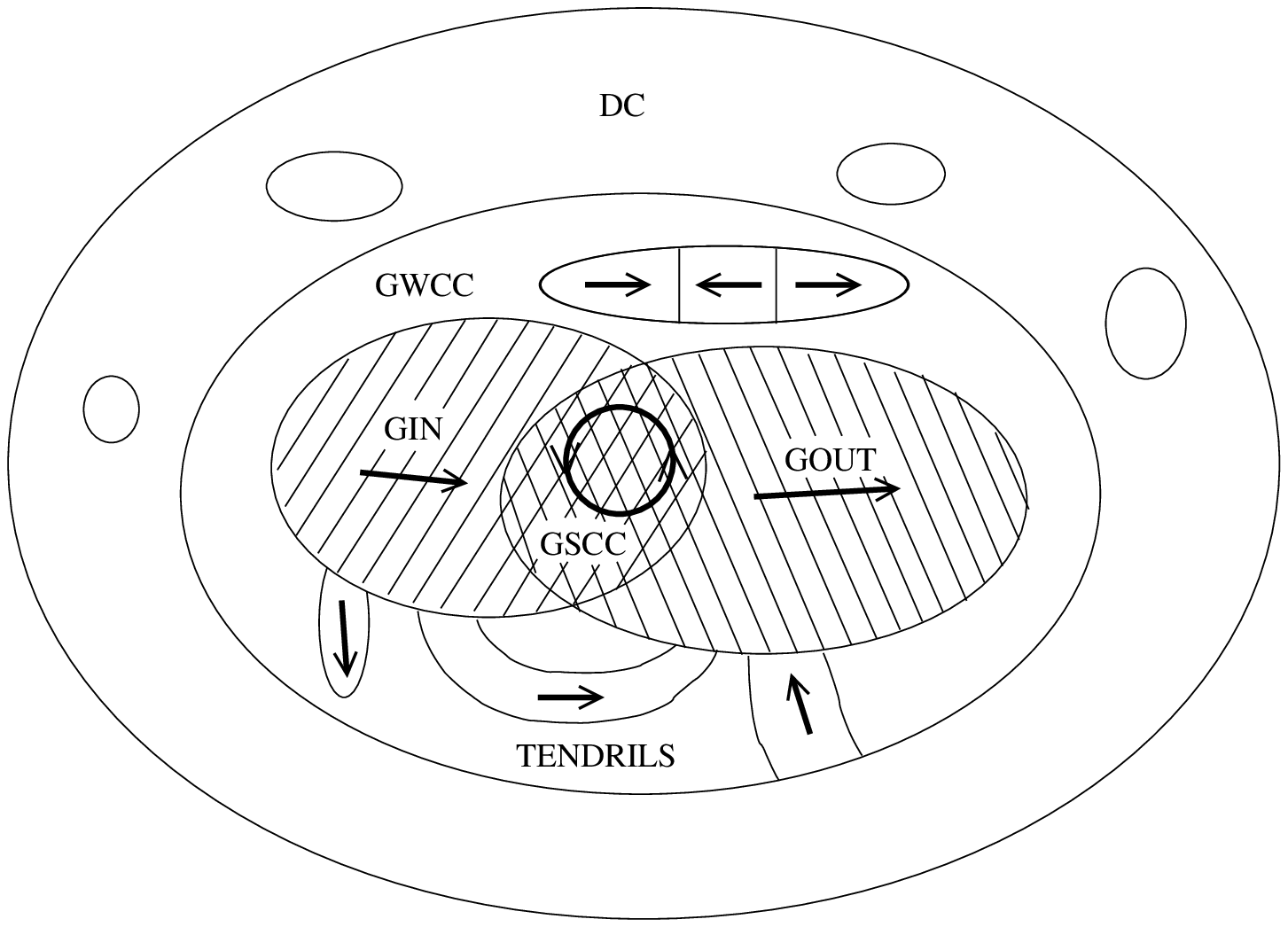}
\caption{
General structure of a directed network in the situation when the giant strongly connected component is present. Also the structure of the WWW (compare with Fig. 9 of Ref. \protect\cite{bkm00}). \\
If one ignores the directedness of edges, the network consists of the {\em giant weakly connected component} (GWCC) -- actually, the usual percolative cluster -- and disconned components (DC). \\ 
Accounting for the directedness of edges, the GWCC contains the following components: \\
(a) the {\em giant strongly connected component} (GSCC), that is the set of vertices reachable from its every vertex by a directed path; \\
(b) the {\em giant out-component} (GOUT), the set of vertices approachable from the GSCC by a directed path; \\
(c) the {\em giant in-component} (GIN), contains all vertices from which the GSCC is approachable; \\
(d) the {\em tendrils}, the rest of the GWCC, i.e., the vertices which have no access to the GSCC and are not reachable from it. In particular, it indeed includes something like ``tendrils'' \protect\cite{bkm00} going out of GIN or coming in the GOUT but also there are ``tubes'' going from the GIN to GOUT without passage through GSCC and numerous clusters which are only ``weakly'' connected. \\
Note that the definitions of the GIN and GOUT in the present paper differ from the definitions of Refs. \protect\cite{bkm00,nsw00}. Here the GSCC is included into both the GIN and GOUT, so the GSCC is the interception of the GIN and GOUT.  
We have to introduce the new definitions for the sake of brevity and logical presentation (see the calculations in the text).  
}
\label{f1}
\end{figure}

The most interesting networks in Nature, including the World Wide Web, are directed graphs, i.e., their vertices are connected by directed edges \cite{nsw00,dms00,dm00c,t01,krr00,er01}. In the general case, the structure of the directed graph looks as it is shown in Fig. \ref{f1} (all the notions are introduced and explained in the figure caption). In particular, 
the World Wide Web has such a structure \cite{bkm00}.

In Refs. \cite{nsw00,cnsw00}, 
the previous strong results of mathematicians \cite{mr95,mr98} were developed, and
it was proposed the general theory of percolation phenomena in networks with arbitrary degree distributions and statistically uncorrelated (randomly connected) vertices. 
%%(i.e. with random connections). 
Of course, the last assumption is not true for most of growing nets in Nature. Nevertheless, the direct conclusions of such an approach proved to explain the behavior of real networks \cite{ajb00}. 

In paper \cite{nsw00}, it was shown how to find the relative sizes of the following giant components of directed graphs with statistically uncorrelated vertices: (i) {\em the giant weakly connected component}, $W$; (ii) {\em the giant in-component}, $I$; and (iii) {\em the giant out-component}, $O$. We emphasize that, for brevity and consistency, we use the definitions of the giant in- and out-components  other than in Refs. \cite{bkm00,nsw00} (see the caption of Fig. \ref{f1}). 

%%In the present communication, 
Here we demonstrate how to calculate the relative size $S$ of, perhaps, the most important part of the directed graph, of {\em its giant strongly connected component} (iv). 
(In the GSSC, every pair of vertices is connected in both directions, 
i.e., from one of the vertices, one can approach the other by moving either along or against the edge directions.)  
This allows us to completely describe the total structure of directed graphs with arbitrary degree distributions and statistically uncorrelated vertices. For the demonstration, we use the networks with the simplest degree distributions providing non-trivial results. 

We have to briefly remind a very usefull approach of Ref. \cite{nsw00}. The Z-transforms (or generating functions) are used. For the undirected graph, 
$\Phi(x) \equiv \sum_{k} P(k) x^k$, and, for the directed one, 
$\Phi(x,y) \equiv \sum_{k_i,k_o} P(k_i,k_o) x^{k_i} y^{k_o}$ \cite{jlrbook00}. 
Here, $P(k) \equiv P^{(w)}(k)=\sum_{k_i}P(k_i,k-k_i)$ is the degree distribution ($k=k_i+k_o$ is the total number of connections of a node) and $ P(k_i,k_o)$ is the joint distribution of in- and out-degrees. 
When all the connections are inside the network, the average in- and out-degrees are equal: $\partial_x\Phi(x,1)\left.\right|_{x=1} = \partial_y\Phi(1,y)\left.\right|_{y=1} \equiv z^{(d)}$. Therefore the average degree is $z=2z^{(d)}$.
If one ignores the directedness of edges, the degree distribution of the directed network, in the Z-representation, takes the form $\Phi^{(w)}(x)=\Phi(x,x)$. 
In this case, the distribution of the number of connections minus one of any of the end vertices of a randomly chosen edge corresponds to 
$\Phi_1^{(w)}(x) \equiv \Phi^{(w)\,\prime}(x)/z$.

\begin{figure}
\epsfxsize=90mm
%%%\epsffile{fig_strong_3.eps}
\epsffile{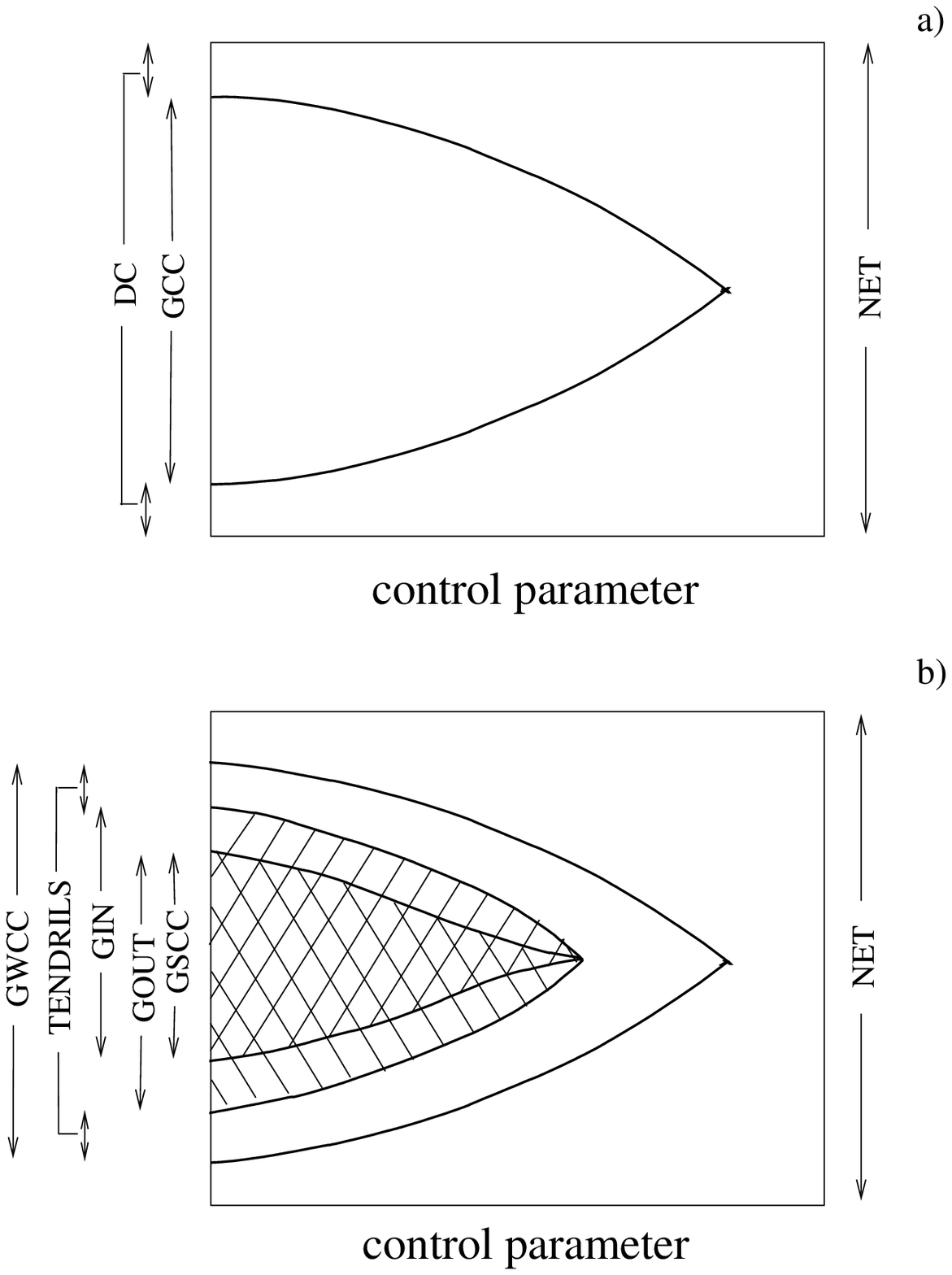}
\caption{
Schematic plots of the variations of all the giant components vs some control parameter for the undirected network $(a)$ and for the directed one $(b)$. In the undirected graph, the meanings of the giant connected component (GCC), i.e., its percolative cluster, and the GWCC coincide.
}
\label{f3}
\end{figure}

The giant weakly connected component exists if $\Phi_1^{(w)\,\prime}(1)>1$, that corresponds to the well known criterium of Molloy and Reed \cite{mr95} 

\begin{equation}
\sum_{k} k(k-2)P(k) > 0
\, .  
\label{1}
\end{equation}   
The size of the GWCC, $W$, can be easily obtained from the relations \cite{nsw00,mr98} 

\begin{equation}
W = 1 - \Phi^{(w)}(t_c) \  , \ \ \ \ x_c = \Phi_1^{(w)}(t_c)
\, .  
\label{2}
\end{equation} 
From Eq. (\ref{1}), one sees that the existence of the GWCC crucially depends on the size of the fraction of dead ends in the network. Indeed, $P(1)$ is the only term in Eq. (\ref{1}) that prevents the GWCC. 
%%For example, in Fig. \ref{f2}, we show the variation of 
%%$W$ of the simplest undirected network with the simplest 
%%degree distribution $P(k) = p\delta_{k,1}+(1-p)\delta_{k,k_0}$ 
%%vs the parameter $p$. The role of the dead ends is clearly visible. 
In Fig. \ref{f3},a, the evolution of the giant connected component of the undirected graph induced by the change of some control parameter is schematically shown.
From Eq. (\ref{1}), it is also clear that the divergency of the second moment of the degree distribution makes the GWCC extremely stable \cite{remark}. 
If the exponent $\gamma$ of the power-law degree distribution $P(k) \sim k^{-\gamma}$ is smaller than $3$ or equal to it, 
%%then, 
%%for the elimination of the GWCC, 
one has to remove at random almost all the vertices or edges of the network to eliminate the GWCC 
%%That is, the GWCC almost surely can not be destroyed by any 
%%random damage if the exponent $\gamma$ of the power-law degree 
%%distribution $P(k) \sim k^{-\gamma}$ is smaller or equal to $3$ 
\cite{cebh00}.  

In a similar way, it is easy to study the GIN and GOUT componets of the directed network \cite{nsw00}. One introduces the Z-transform of the out-degree distribution of the vertex approachable by following a randomly chosen edge when one moves 
{\em along} the edge direction, $\Phi_1^{(o)}(y) \equiv \partial_x\Phi(x,y)\left.\right|_{x=1}/z^{(d)}$. Also, 
$\Phi_1^{(i)}(x) \equiv \partial_y\Phi(x,y)\left.\right|_{y=1}/z^{(d)}$ corresponds to the in-degree distribution of the vertex which one can approach moving {\em against} the edge direction. The GIN and GOUT are present if 
$\Phi_1^{(i)\,\prime}(1) = \Phi_1^{(o)\,\prime}(1) = \partial^2_{xy}\Phi(x,y)\left.\right|_{x=1,y=1}/z^{(d)} > 1$, that is \cite{nsw00} 

\begin{eqnarray}
& & \sum_{k_i,k_o} (2 k_i k_o - k_i - k_o)P(k_i,k_o) = 
\nonumber
\\[5pt]
& & 2\! \sum_{k_i,k_o} \! k_i(k_o - 1)P(k_i,k_o) =
2\! \sum_{k_i,k_o} \! k_o(k_i - 1)P(k_i,k_o) > 0 
\, .
\nonumber
\\
& &  
\label{3}
\end{eqnarray}
In this case, there exist the non-trivial solutions of the equations 

\begin{equation}
x_c = \Phi_1^{(i)}(x_c) \  , \ \ \ \ y_c = \Phi_1^{(o)}(y_c)
\, .  
\label{4}
\end{equation} 
They have the following meanings. 
$x_c<1$ is the probability that the connected component, obtained by moving {\em against} the edge directions starting from a randomly chosen edge, is finite. 
$y_c<1$ is the probability that the connected component, obtained by moving {\em along} the edge directions starting from a randomly chosen edge, is finite. 
Then $P(k_i,k_o)x_c^{k_i}$ and  $P(k_i,k_o)y_c^{k_o}$ are the probabilities that a vertex with $k_i$ incoming and $k_o$ outgoing
edges has finite in- and out-components, respectively. 
The in- and out-components of a vertex are sets of vertices which are approachable from it moving against and along edges, respectively, plus the vertex itself.  
Summation of these expressions over $(k_i,k_o)$ yields the total probabilities that the in- and out-components of a randomly chosen vertex are finite, respectively. 
Therefore, they are equal to $\Phi(x_c,1)$ and $\Phi(1,y_c)$, respectively. 
Thus, the relative sizes of the GIN and GOUT are
%%%from which one can get the sizes of the GIN and GOUT,  

\begin{equation}
I = 1 - \Phi(x_c,1) \  , \ \ \ \ O = 1 - \Phi(1,y_c)
\, .  
\label{5}
\end{equation} 

Here we show that from Eq. (\ref{4}), it is possible to find not only $I$ and $O$ but also the relative size $S$ of the GSCC using the considerations similar to Ref. \cite{nsw00}. 
Suppose that a vertex has $k_i$ incoming and $k_o$ outgoing edges. 
They are assumed to be statistically independent. 
Then the probability that all the incoming edges come from finite in-components is $x_c^{k_i}$. 
The probability that this vertex has the infinite in-component is equal to $1-x_c^{k_i}$, that is, at least one of the $k_i$ incoming edges has to come from the GIN. 
Similarly, $1-y_c^{k_o}$ is the probability that the vertex has the infinite out-component. The vertex belongs to the GSCC if its in- and out-components are both infinite; the corresponding probability is equal to 
$(1-x_c^{k_i})(1-y_c^{k_o})$. Then the total probability that a vertex belongs to the GSCC is equal to 
$\sum_{k_i,k_o} P(k_i,k_o)(1 - x_c^{k_i})(1 - y_c^{k_o})$.
Finally, the relative size of the GSCC takes the form 
%%%% 
%%%%
%%%%Let us show that from the solutions of Eq. (\ref{4}), 
%%%%it is possible to find not only $I$ and $O$ but also the 
%%%%relative size $S$ of the GSCC. 
%%%%The values $\Phi(x_c,1)$, $\Phi(1,y_c)$, and $\Phi(x_c,y_c)$ 
%%%%have the following meaning. 
%%%%Actually, $\Phi(x_c,1)$ is the probability that a vertex has 
%%%%a finite in-component. 
%%%%The in-component of a vertex is the set of vertices which 
%%%%are approachable from it plus the vertex itself.  
%%%%$\Phi(1,y_c)$ is the probability that a vertex has a finite out-component. 
%%%%The out-component of a vertex is the set of vertices from which 
%%%%it can be reached plus the vertex itself.   
%%%%$\Phi(x_c,y_c)$ is equal to the probability that both the 
%%%%in- and out-components of a vertex are finite 
%%%%(we use the considerations similar to Ref. \cite{nsw00}). 
%%%%The size of the GSCC, $S$, equals the probability that a 
%%%%randomly chosen vertex has an infinite in- and an infinite 
%%%%out-components. Therefore, 

\begin{eqnarray}
S & = &
\sum_{k_i,k_o} P(k_i,k_o)(1 - x_c^{k_i})(1 - y_c^{k_o})
%%1 - \Phi(x_c,1) - \Phi(1,y_c) + \Phi(x_c,y_c) 
=
\nonumber
\\[5pt]
& & 
%%\sum_{k_i,k_o} P(k_i,k_o)(1 - x_c^{k_i})(1 - y_c^{k_o}) 
1 - \Phi(x_c,1) - \Phi(1,y_c) + \Phi(x_c,y_c)
\, .
\label{6}
\end{eqnarray} 
Therefore, $\Phi(x_c,y_c)$ is equal to the probability that both the in- and out-components of a vertex are finite. 
One can write $\Phi(x_c,y_c) = 1-W+T$.
Knowing $W$, $S$, $I$, and $O$, it is easy to obtain the relative size of 
{\em tendrils},
%%``TENDRILS'',

\begin{equation}
T = W + S - I - O
\, .  
\label{7}
\end{equation} 
%%One can also write $\Phi(x_c,y_c) = 1-W+T$. 

Eqs. (\ref{2}), (\ref{4})--(\ref{7}) allow us to obtain all the giant components of the directed networks with arbitrary joint in,out-degree distributions and statistically uncorrelated vertices. It is usefull to rewrite the main Eq. (\ref{6}) in the form

\begin{equation}
S = I O + \Phi(x_c,y_c) - \Phi(x_c,1)\Phi(1,y_c)
\, .  
\label{8}
\end{equation} 
If the the joint distribution of in- and out-degrees factorizes, 
$P(k_i,k_o)=P^{(i)}(k_i)P^{(o)}(k_o)$, Eq. (\ref{8}) takes the simple form $S=IO$. 
Otherwise, such factorization of $S$ is impossible. At the threshold, $x_c=y_c=1$, and $I$, $O$, and $S$ simultaneously approach zero.  

\begin{figure}
\epsfxsize=80mm
%%%\epsffile{fig_strong_4.eps}
\epsffile{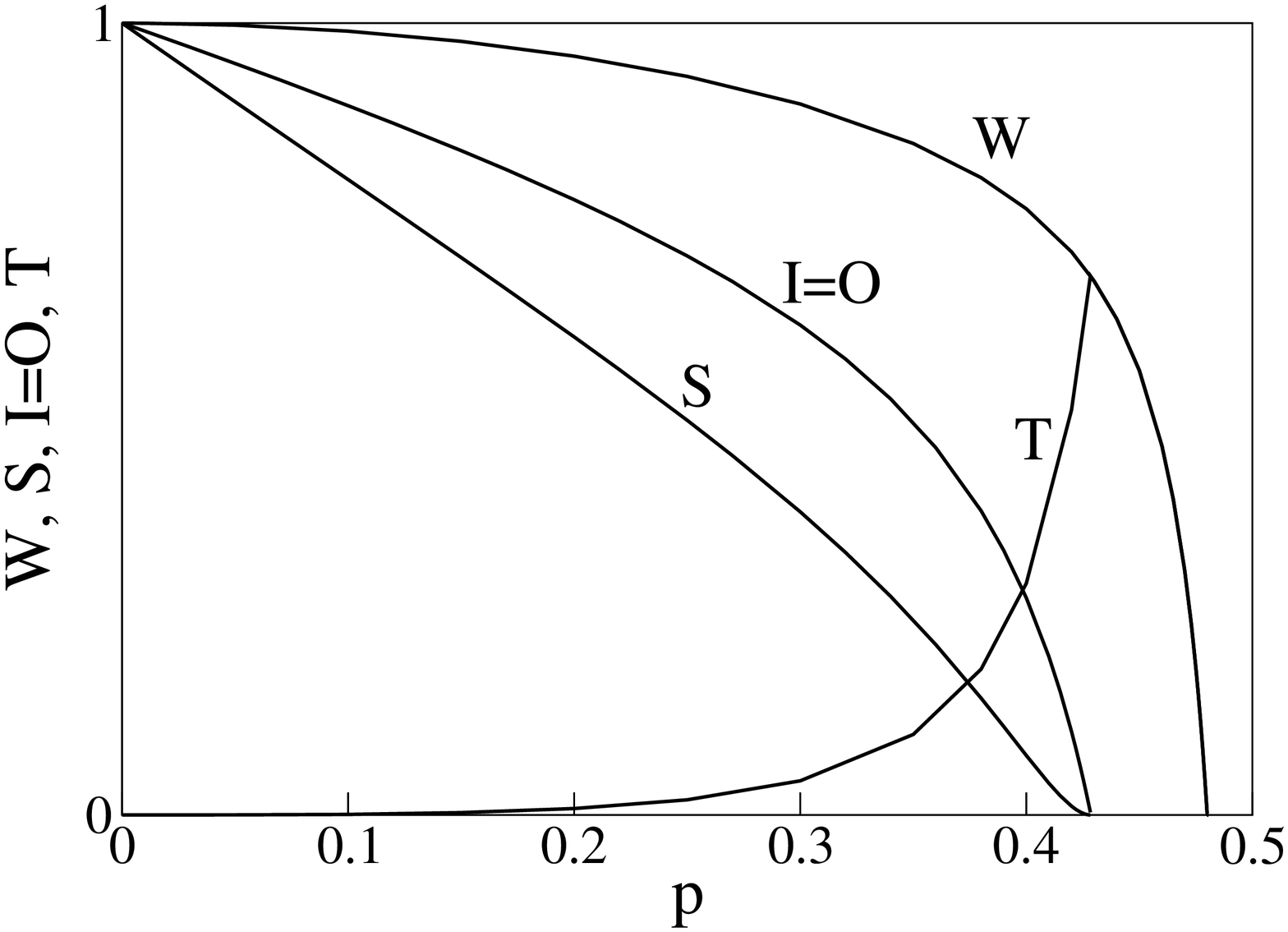}
\caption{
Relative sizes of the GWCC $(W)$, GSCC $(S)$, GIN $(I)$, GOUT $(O)$, and TENDRILS $(T)$ vs the parameter $p$ for the directed graph with the factorizable joint in,out-degree distribution 
$P(k_i,k_o)=[p(\delta_{k_i,0}+\delta_{k_i,1})+(1-2p)\delta_{k_i,3}]\times$ 
$[p(\delta_{k_o,0}+\delta_{k_o,1})+(1-2p)\delta_{k_o,3}]$. 
In this case, $S=IO$, $I=O$.
}
\label{f4}
\end{figure}

\begin{figure}
\epsfxsize=80mm
%%%\epsffile{fig_strong_5.eps} 
\epsffile{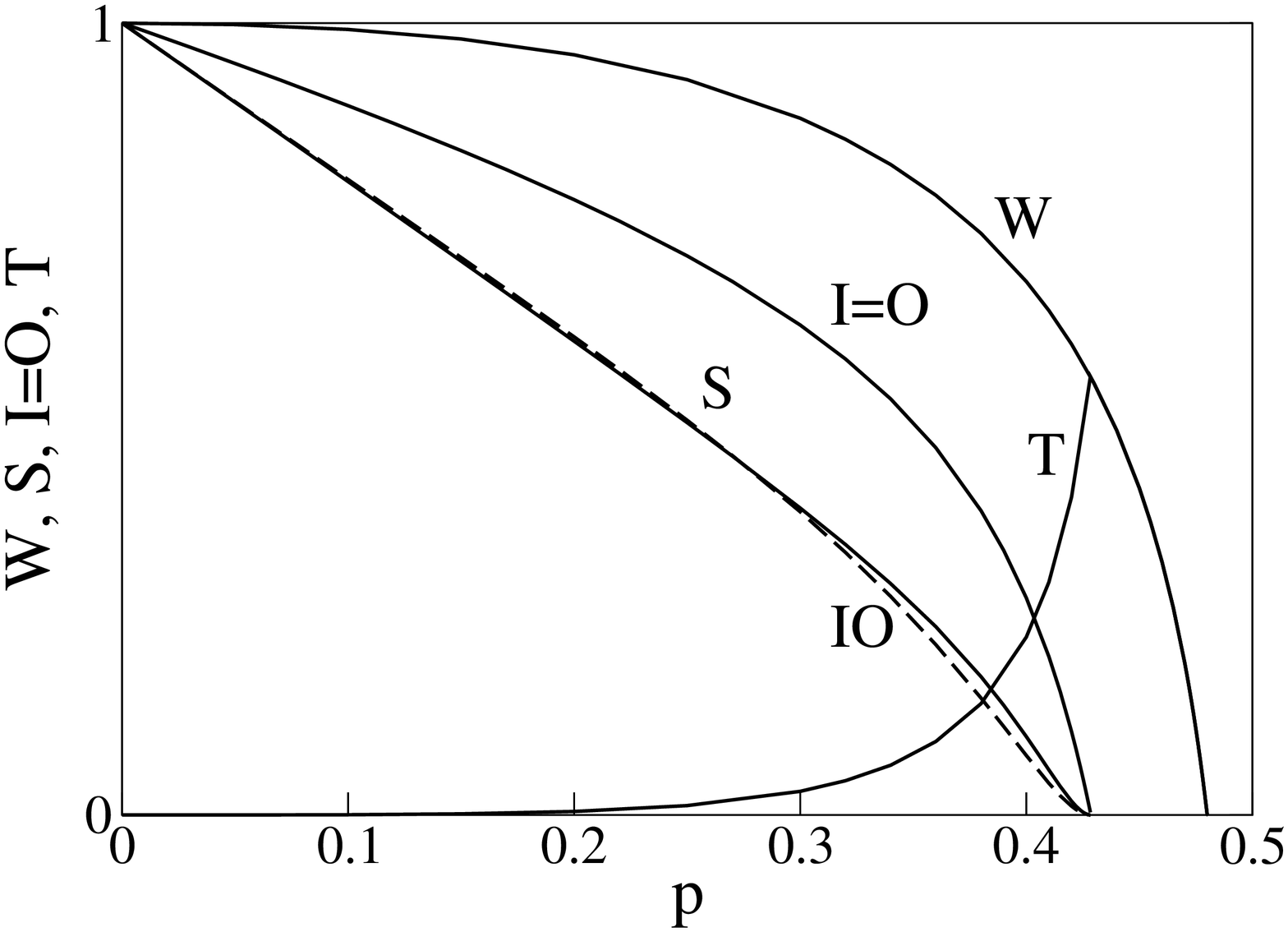}
\caption{
Relative sizes of the GWCC $(W)$, GSCC $(S)$, GIN $(I)$, GOUT $(O)$, and TENDRILS $(T)$ vs the parameter $p$ for the directed graph with the joint in,out-degree distribution 
$P(k_i,k_o) = p(\delta_{k_i,0}\delta_{k_o,1} + \delta_{k_i,1}\delta_{k_o,0}) + (1-2p)\delta_{k_i,3}\delta_{k_o,3}$ that does not factorize. This form of the distribution means that if a node is of large in-degree, then its out-degree is also large. Also, if the degree of a node is small, the out-degree is small too.  
The dashed curve shows the product $IO$ (compare with the curve for $S$). In the particular case that we consider here, $I=O$.
}
\label{f5}
\end{figure}

We have no intention to calculate the sizes of the giant components for real networks, for instance, for the WWW, for the following reasons.  
There are some correlations between their vertices, and their joint degree distributions are unknown yet (nevertheless, see the attempt of the calculation of $I$ and $O$ for the WWW in Ref. \cite{nsw00}). 
Instead of this, for demonstration, we consider two simplest non-trivial nets. In the first of them, the joint in,out-degree distribution factorizes: $P(k_i,k_o)=[p(\delta_{k_i,0}+\delta_{k_i,1})+(1-2p)\delta_{k_i,3}]\times$ 
$[p(\delta_{k_o,0}+\delta_{k_o,1})+(1-2p)\delta_{k_o,3}]$. The results, the dependence  of the sizes of the giant components on $p$ are shown in Fig. \ref{f4} and, schematically, in Fig. \ref{f3},b. The curves $I(p)=O(p)$ approach the threshold linearly, and $S(p)$ -- quadratically but the range of the quadratic dependence is narrow. 
Over a wide range of $p$, $S(p) \approx 1/2-p$ (also see the next case, Fig. \ref{f5}).

In real growing nets, the joint in,out-degree distributions do not factorize just because of their growth \cite{krr00,krl00,kr00}. Therefore, for comparison, we calculate the sizes of the giant components for the network with the distribution  
$P(k_i,k_o) = p(\delta_{k_i,0}\delta_{k_o,1} + \delta_{k_i,1}\delta_{k_o,0}) + (1-2p)\delta_{k_i,3}\delta_{k_o,3}$. This means that large in- and out-degrees correlate as well as small in- and out-degrees. The results are plotted in Fig. \ref{f5} (also see the schematic plot \ref{f3},b) \cite{remark2}. One sees that, in this case, the size of the GSCC noticeably differs from the product $IO$. We should note that a similar deviation is present in the WWW. From the data of Ref. \cite{bkm00} for the WWW, $I \approx 0.490$, $O \approx 0.489$, so $IO \approx 0.240$, that is less than the measured value $S \approx 0.277$ but is not far from it. 

%%%Note that even if the average degree $z$ is large, 
%%%the size $W$ of the GWCC, in principle, may subsequently deviate from $1$. 
%%%One can easily check this using, e.g., degree distributions similar 
%%%to the distributions considered above. 

%%If the average degree $z$ is large, the size $W$ of the GWCC 
%%is close to $1$. Nevertheless, the resulting value is strongly 
%%dependent on the particular forms of degree distributions. For 
%%instance, is $z=15$ as for the WWW \cite{bkm00}, then 
%%(i) the Poisson degree distribution provides 
%%$W \cong 1-e^{-z} \approx 1- 3\times 10^{-7}$ \cite{bbook85}, 
%%(ii) the factorizable distribution considered above with the substitution 
%%$\delta_{k_{i,o},3} \to \delta_{k_{i,o},15}$ and $p=0.357$ 
%%gives $W = 0.868$, 
%%and  
%%(iii) the non-factorizable distribution discussed above with the 
%%same replacement and $p$ provides $W=0.966$. 

Figures \ref{f3}--\ref{f5} demonstrate that, in the wide enough range of parameters, the following situation may be realized. The directed graph may have the GWCC and have no the GSCC. Only the stability of the GWCC to random damage was discussed yet \cite{cebh00}. Nevertheless, just the stability of the GSCC is the most important, e.g., for the WWW.
Is it possible that the GSCC are less resilient to failures than the GWCC? Let us briefly discuss this problem. 

One can see from Eq. (\ref{3}), that the GSCC is extremely resilient if the average $\langle k_ik_o \rangle$ diverges. Let us consider two limiting situations. In the first one, the joint in,out-degree distribution factorizes, so 
$\langle k_ik_o \rangle = \langle k_i \rangle^2$. In this case, if the distributions are of a power-law form then, for the robustness of the GSCC, the corresponding exponent $\gamma_{i}$ or $\gamma_{o}$ should be smaller than $2$ or equal it. This is a very strong requirement.
Here, the smallest exponent of $\gamma_{i}$ and $\gamma_{o}$ is also equal to the exponent $\gamma$ of the degree distribution. 
For the resilience of the GWCC to random damage, it should be $\gamma \leq 3$ \cite{cebh00} that ensures the divergence of $\langle k^2 \rangle$. Therefore, for such distributions, when $2 < \min(\gamma_{i},\gamma_{o})=\gamma \leq 3$, random damage can destroy the GSCC but can not eliminate the GWCC. 

In the other limiting case, $P(k_i,k_o) = P(k_i)\delta_{k_i,k_o}$, the correlations between in- and out-degrees are very strong. This form more resembles the joint degree distributions of the real growing networks. In such an event, $\langle k_i k_o \rangle = \langle k_i^2 \rangle$, and the conditions for the resilience of the GSCC and GWCC, $\gamma \leq 3$, coincide. One should note that the real distributions are between the considered limiting cases.

In summary, we have shown how to obtain the size of the giant strongly connected component of the directed network with the arbitrary degree distribution and statistically uncorelated vertices. This allows us to find all the giant components of such a graph and to describe its basic structure. Using the simplest examples and the general considerations we have demonstrated that the correlations between in- and out-degrees subsequently influence the global topology of the network. 
\\

SND thanks PRAXIS XXI (Portugal) for a research grant PRAXIS XXI/BCC/16418/98. SND and JFFM 
were partially supported by the project POCTI/1999/FIS/33141. We also thank P.L. Krapivsky for useful discussions. 
\\

\noindent
$^{\ast}$      Electronic address: sdorogov@fc.up.pt \\
$^{\dagger}$   Electronic address: jfmendes@fc.up.pt \\
$^{\ddag }$ Electronic address: alnis@samaln.ioffe.rssi.ru

\newpage

%%\newpage
%%
%%\begin{figure}
%%\epsfxsize=80mm
%%\epsffile{fig_strong_2.eps}
%%\caption{ 
%%Relative size $W$ of the giant connected component vs 
%%parameter $p$ for the undirected network with the degree 
%%distribution $P(k) = p\delta_{k,1}+(1-p)\delta_{k,k_0}$. 
%%Here, $p$ is the probability that a node is a dead end. 
%%The curves correspond to the following values of $k_0$: 
%%$3$, $6$, and $30$.  
%%}
%%\label{f2}
%%\end{figure}

%%\newpage

%%\hspace{4cm}
\newpage

\hspace{10cm}
%%\newpage

\end{multicols}


\begin{references}
\bibitem{remark} Rigorously speaking, to study the resilience of the giant connected components to random damage, one has to consider networks with randomly deleted vertices or edges (see Refs. \protect\cite{cnsw00} and \protect\cite{cebh00}). Nevertheless, the divergence of the moments of the undamaged network already indicates its anomalous resilience, since it cannot be removed by random damage \protect\cite{cnsw00,cebh00}.
\bibitem{remark2} Note that even if the average degree $z$ is large, the size $W$ of the GWCC, in principle, may subsequently deviate from $1$. One can easily check this using, e.g., degree distributions similar to the distributions considered above. 
The Poisson distribution with large $z$ produces extremely small values  
$1-W \cong 
%%e^{-z}$
\exp[-z]$ 
since there are few dead ends in this case.
\end{references}
\end{document}